\documentclass[reprint,superscriptaddress,amsmath,amssymb,aps,physrev]{revtex4-2}

\usepackage{graphicx}
\usepackage{dcolumn}
\usepackage{bm}

\newcommand{\omegap}{\omega_\perp}
\newcommand{\gOneD}{g_\textrm{1D}}

\newcommand{\aOneD}{a_\textrm{1D}}
\newcommand{\nOneD}{n_\textrm{1D}}
\newcommand{\aThreeD}{a_\textrm{3D}}
\newcommand{\nThreeD}{n_\textrm{3D}}
\newcommand{\Er}{E_\textrm{r}}
\newcommand{\kB}{k_{\textrm{\tiny B}}}
\newcommand{\RTF}{R_{\textrm{TF}}}
\newcommand{\lx}{\ell_x}
\newcommand{\xiT}{\xi_{T}}
\newcommand{\xig}{\xi_{\gamma}}
\newcommand{\kc}{k_{\textrm{c}}}
\newcommand{\dif}{\textrm{d}}

\begin{document}

\preprint{APS/123-QED}

\title{\textbf{Direct measurement of Tan's contact in a one-dimensional Lieb-Liniger gas} 
}%

\author{Qi Huang}
\affiliation{Institute of Quantum Electronics, School of Electronics, Peking University, No.5 Yiheyuan Road, Haidian District, Beijing, 100871, China.}
\affiliation{International Center for Quantum Materials, School of Physics, Peking University, No.5 Yiheyuan Road, Haidian District, Beijing, 100871, China.}

\author{Hepeng Yao}
\email{Hepeng.Yao@unige.ch}
\affiliation{DQMP, University of Geneva, 24 Quai Ernest-Ansermet, Geneva, CH-1211, Switzerland.}

\author{Xuzong Chen}
\email{xuzongchen@pku.edu.cn}
\affiliation{Institute of Quantum Electronics, School of Electronics, Peking University, No.5 Yiheyuan Road, Haidian District, Beijing, 100871, China.}

\author{Laurent Sanchez-Palencia}
\email{laurent.sanchez-palencia@polytechnique.edu}
\affiliation{CPHT, CNRS, Ecole Polytechnique, IP Paris, 91120 Palaiseau, France.}

\begin{abstract}
The Tan contact has emerged as a pivotal quantity in characterizing many-body quantum systems, bridging microscopic short-range correlations to thermodynamic behavior. It is defined as the weight of universal $1/k^4$ fall off in momentum distribution tails, which can be measured directly in ultracold gases. So far, however, its direct measurement has been hindered in Bose gases due to interactions strongly affecting the expansion dynamics. Here, we present the first direct measurement of the Tan contact in a strongly-correlated Lieb-Liniger gas. Leveraging the one-dimensional geometry of our system, we implement a two-stage expansion scheme, yielding interaction-immune time-of-flight imaging. Our results show excellent agreement with theoretical predictions from quantum Monte Carlo calculations, which also provides independent thermometry of the experiment. By varying atom number, temperature, and interaction strength, we obtain results consistent with the universal scaling law predicted for the trapped Lieb-Liniger model. Our work paves the way for further characterization of the Lieb-Liniger gas across broad interaction regimes and holds promise for extension to other correlated quantum gases in confined geometries.
\end{abstract}

\maketitle

Interactions play a pivotal role in determining the physical behaviour of many-body systems and drive quantum phase transitions~\cite{sachdev2011,bruus2004,tsvelik2007}.
The simplest model, which considers pairwise contact interactions, is sufficient to account for the main features of most many-body systems in condensed matter and strictly applies to most ultracold atomic gases, where the interaction strength can also be tuned via magnetic Feshbach resonances~\cite{lewenstein2007,bloch2008,chin2010}. Such contact interaction implies a zero-distance singularity of the many-body wavefunction, which manifests in momentum space as a characteristic momentum distribution with large-momentum tails scaling as $n(k) \simeq C/k^4$~\cite{korepin1997,olshanii2003,tan2008_3}. This universal behaviour applies irrespective of the nature of particles, system dimensionality, temperature, and interaction strength. The constant $C$, known as the Tan contact, may be interpreted as a thermodynamic conjugate of the interaction strength~\cite{tan2008_3,zou2021} and is related to a number of quantities,
including thermodynamic potentials, pressure, entropy, as well as interaction energy, and two-body correlations, within the so-called Tan relations~\cite{tan2008_1,tan2008_2}. In one dimension (1D), recent theoretical work has shown that the Tan contact provides fruitful information about the specific effects of strong correlations, including the celebrated Tonks-Girardeau fermionization of strongly-interacting Bose gases~\cite{barth2011,minguzzi2002,vignolo2013,chen2014,xu2015,yao2018,decamp2018,patu2018}.

For most ultracold atomic gases, pairwise contact interaction is guaranteed by dilution and s-wave scattering at low energy~\cite{lewenstein2007,bloch2008}, with noticeable exceptions for dipolar and Rydberg-atom gases. In recent years, several measurements of the Tan contact have been reported for interacting Fermi and Bose gases~\cite{stewart2010,kuhnle2010,wild2012,laurent2017,mukherjee2019,carcy2019,zou2021}. Strongly-interacting Fermi gases have allowed to measure the contact by various techniques~\cite{stewart2010,kuhnle2010,mukherjee2019,carcy2019}, and verify universal Tan's relations~\cite{stewart2010}. The Tan contact has also been measured for weakly-interacting Bose gases in two (2D) and three (3D) dimensional systems, using rf spectroscopy~\cite{wild2012} and Ramsey interferometry~\cite{zou2021}. In contrast, exploration of the Tan contact for strongly-interacting Bose gases, in particular in 1D, is still missing. So far a major impediment to direct observation of universal $1/k^4$ momentum tails and direct measurement of the contact is that interactions during standard time of flight (TOF) strongly affect the momentum distribution~\cite{qu2016}. Moreover, impurities, as created by spin-flip processes in magnetic traps, have been shown to also alter the expansion dynamics and produce dynamical $1/k^4$ tails with a strongly modified contact~\cite{chang2016,cayla2023}.

In this work, we report the first observation of universal $1/k^4$ tails in the momentum distribution of a strongly-interacting 1D Bose (Lieb-Liniger) gas, and direct measurement of the Tan contact from their weight. Using a purely optical confinement and leveraging the 1D geometry, we realize a contact-preserving TOF measurement of the momentum distribution via a two-stage expansion scheme. The experimental data for the momentum distributions and the measured contacts show good agreement with quantum Monte Carlo (QMC) calculations, for various values of temperature and particle number. Our results show clear beyond-mean field many-body effects and provide the first verification of the predicted universal two-parameter scaling of the contact for trapped, finite-temperature, Lieb-Liniger gases~\cite{yao2018}.

\section{Emulating the trapped Lieb-Liniger model}

The extended Lieb-Liniger model we consider is governed by the Hamiltonian
\begin{equation}\label{eq:H}
	\hat{H}=\sum_{i}\left(-\frac{\hbar^{2}}{2 m} \frac{\partial^{2}}{\partial x_{i}^{2}}+V\left(x_{i}\right)\right)+\gOneD \sum_{i<j} \delta\left(x_{i}-x_{j}\right),
\end{equation}
where $\hbar$ is the reduced Planck constant, $m$ is the atomic mass, $x_i$ is the position of particle $i$, and $V\left(x\right)=m \omega_x^2 x^2/2$ is an external harmonic potential with frequency $\omega_x$. The second sum accounts for point-like pairwise interactions and the coupling constant may be related to the 1D scattering length $\aOneD$ via the formula $\gOneD = -2\hbar^2/m \aOneD$~\cite{olshanii1998,petrov2000}. In ultracold atom systems, it is realized by strongly confining a 3D Bose gas to zero-point transverse oscillations using a strong transverse harmonic trap of frequency $\omega_\perp$. The 1D scattering length $\aOneD$ is then related to the 3D scattering length $\aThreeD$ and the transverse oscillation length $\ell_\perp = \sqrt{\hbar/m \omega_\perp}$, via $\aOneD = -\ell_\perp \left(\ell_\perp/\aThreeD - C_0 \right)$ with $C_0=\lvert \zeta(1/2) \rvert/\sqrt{2}\approx 1.0326 $ and $\zeta$ the Riemann zeta function~\cite{olshanii1998,bloch2008}.  For an homogeneous gas with 1D density $\nOneD$, the interaction regime is characterized by the Lieb-Liniger parameter $\gamma=m\gOneD/\hbar^2 \nOneD$~\cite{lieb1963a,lieb1963b}.
For $\gamma \ll1$ (high density), the system is weakly interacting and may be approximately described using mean field theory. For stronger interactions, $\gamma \gg 1$ (low density), the Bose gas crosses over toward the Tonks-Girardeau (TG) regime, where the bosonic many-body wavefunction can be mapped onto that of free fermions. This effect is known as Tonks-Girardeau fermionization~\cite{girardeau1960}.

\begin{figure}[ht]
	\centering
	\includegraphics[width=8.6cm]{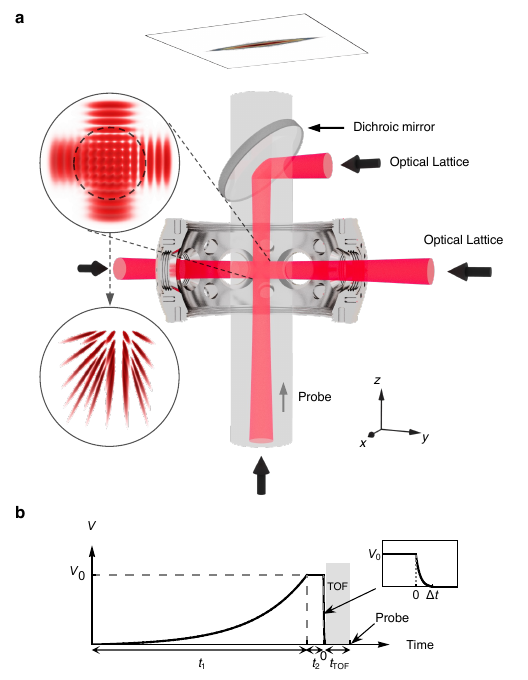}
	\caption{
		\textbf{Sketch of the experimental setup and sequence.}
		\textbf{a},~An array of one-dimensional tubes are generated by a strong 2D lattice potential realized by the standing wave from two retro-reflected laser beams along the axes $y$ and $z$ (red). The tubes are then formed along the $x$ direction (insets) and imaging is performed along the $z$ direction.
		\textbf{b},~The 2D lattice is progressively ramped up to from $0$ to $V_0$ over a time $t_1=260\,\mathrm{ms}$ and it is then held for a further $t_2=20\,\mathrm{ms}$. Time-of-flight expansion is induced by ramping down the lattice beams to zero in $\Delta t = 50\,\mu\mathrm{s}$ and let to expand freely for another  $\Delta t = 30\,\mathrm{ms}$.
	}\label{fig1}
\end{figure}

In our experiment, we start with a 3D Bose-Einstein condensate (BEC) of $^{87}\textrm{Rb}$ atoms confined in an optical dipole trap, with minimum temperature less than $20\,\textrm{nK}$. The atom number of the nearly pure BEC varies from $3\times10^4$ to $1.5\times10^5$, depending on the chosen experimental parameters. We then load the BEC into a strong 2D optical lattice formed by an orthogonal pair of retro-reflected laser beams with wavelength $1064\,\textrm{nm}$ in the $y$ and $z$ directions~(Fig.\,\ref{fig1}a). The loading process is realized by ramping up exponentially the intensity of the optical lattice from $0$ to $V_0=70\,\Er$ in a time $t_1 = 260\,\textrm{ms}$~(Fig.\,\ref{fig1}b). Here $\Er=\hbar^2k_{\textrm{\tiny L}}^2/2m$ is the recoil energy with $k_{\textrm{\tiny L}}$ the laser wave vector.
The 2D optical lattice generates an array of independent, parallel 1D tubes orthogonal to the laser beams~(see insets of Fig.\,\ref{fig1}a). For such a large laser intensity, tunnel coupling between the tubes is negligible, as evidenced by the absence of observable interference in 3D TOF imaging. We then hold the system for a further $t_2 = 20\,\textrm{ms}$ and let it equilibrate. For a 2D lattice amplitude of $V_0=70\Er$, the transverse confinement is nearly harmonic, with trapping frequency $\omega_\perp/2\pi \approx 33.9$\,kHz. After loading in the 1D tubes, the maximum temperature is $T \simeq 30\, \textrm{nK}$ ($\kB T/2\pi\hbar \simeq 0.625$\,kHz) and the maximum chemical potential is $\mu/2\pi\hbar \simeq 1.7$\,kHz, as estimated by comparison to QMC calculations (see below). It very well satisfies the quasi-1D condition, $\kB T, \mu \ll \hbar\omega_\perp$, with $\kB$ the Boltzmann constant, for all results presented herebelow. Moreover, on top of the strong transverse confinement, the Gaussian-shaped lattice laser beams create an axial harmonic confinement along the tubes with frequency $\omega_x/2\pi \approx 84\,\textrm{Hz}$.  For $^{87}$Rb atoms ($\aThreeD \approx 5.3\,\mathrm{nm}$), it yields to $\aOneD\approx -5.8 \times 10^{-7}\,\mathrm{m}$ and $\gamma$ in the range from $1$ to $1.4$. Except whenever mentioned, we use these parameters in the following. 

\section{Two-stage expansion scheme}

To realize a TOF insensitive to interactions in the axial direction $x$, we take advantage of the quasi-1D structure of the tubes. Switching off the confinement laser beams instantaneously would induce very fast expansion in the transverse ($y$ and $z$) directions, driven by the transverse-confinement kinetic energy in a characteristic time $t_\perp \sim 1/\omega_\perp \simeq 5\, \mathrm{\mu s}$, and a much slower expansion in the axial direction ($x$), with a characteristic time $t_x \sim 1/\omega_x \simeq 2\,\mathrm{m s}$. The transverse expansion by a factor $b(t)>1$ is associated with a sharp reduction of the effective 1D coupling constant $\gOneD^{\mathrm{eff}}=\gOneD/b(t)^2$, so that the effect of interactions along $x$ becomes negligible during the axial expansion stage. In practice, we use a controlled exponential ramp to switch of the confinement progressively in a time $\Delta t$ and then let the gas expand for a longer time $t_{\textrm{TOF}}$~(see Fig.\,\ref{fig1}b). Setting the ramping time such that $t_\perp \ll \Delta t \ll t_x$ ensures that the transverse expansion is slowed down while the switch-off remains nearly instantaneous as regards the axial expansion. In the experiment, we use $\Delta t = 50\,\mathrm{\mu s}$, close to the geometric average of the transverse and axial characteristic times $t_\perp$ and $t_x$.

It remains to make sure that the transverse dilution is sufficient when the axial expansion starts, so that the latter is not significantly affected by interactions.
Owing to the strong initial transverse confinement, the transverse expansion is that of a weakly-interacting, 2D Bose gas, and may be treated in the mean field approximation~(see details and a complete derivation in Ref.~\cite{NoteSuppl}).
For harmonic confinement, this expansion is strictly self-similar in 2D with a time-dependent expansion factor $b(t)$, solution of the second-order differential equation $\ddot{b}+\omegap^2(t) b = {\omegap^2(0)}/{b^3}$~\cite{castin1996,kagan1996}, where $\omegap(t)$ is the time-dependent transverse harmonic trap frequency during ($0<t<\Delta t$ and $\omegap(t) > 0$) and after ($\Delta t< t < \Delta t + t_{\textrm{TOF}}$ and $\omegap(t)=0$) the ramp. The transverse expansion does not affect the 1D density for it results from integration over the transverse directions, $\nOneD(x) = \int \dif y \dif z\, \nThreeD(x,y,z)$. In contrast, after an expansion duration $t$, with $0< t < \Delta t + t_{\textrm{TOF}}$, the effective 1D coupling constant has decreased by a factor of $1/b(t)^2$, $\gOneD(t) = \gOneD/b(t)^2$. For $b(t)$ sufficiently large, the axial expansion stage is now that of a 1D gas in the weakly-interacting regime. The mean field interaction term is $\gOneD(t) \nOneD$ and affects the expansion up to a cut-off momentum $\kc$ such that $\gOneD \nOneD / b^2(1/10\omega_x) \lesssim \hbar^2 \kc^2/2m$, where we set a conservative starting time of the axial expansion at $t=1/10\omega_x \ll 1/\omega_x$. The universal $1/k^4$ tails are expected for $k \gtrsim 1/\vert\aOneD\vert$~\cite{tan2008_1,xu2015}. Using the expression $\gOneD=-2\hbar^2/m\aOneD$ and replacing $\kc$ by $1/\vert\aOneD\vert$, we obtain the criterion
\begin{equation}\label{eq:deltat criteria2}
	4 \vert\aOneD\vert \nOneD \lesssim b(1/10\omega_x)^2,
\end{equation}
such that the $1/k^4$ tails are not significantly affected by the residual interactions. Solving the equation for $b(t)$ for the exponential ramp of the transverse confinement $\omega_\perp(t)$ used in the experiment, we checked that the condition~(\ref{eq:deltat criteria2}) is satisfied for all results presented here. Specifically, we find $5.2 \lesssim 4 \vert\aOneD\vert \nOneD \lesssim 8$
and $b(1/10\omega_x) \approx 17.8$ in our experiments, which sufficiently fulfills the condition~\cite{NoteSuppl}.

Moreover, our system benefits from using all-optical evaporative cooling.
It helps us avoid the influence of impurities created by a magnetic trap during evaporative cooling, which can significantly affect the amplitude of the $1/k^4$ tails~\cite{cayla2023}. Finally, our experimental setup allows us to perform a TOF of $t_{\textrm{\tiny TOF}}=30$\,ms. For the $^{87}$Rb atoms used in the experiment and the initial length of the tubes, $L \lesssim 20\,\mu\text{m}$, it satisfies the far-field condition $t_{\textrm{TOF}} \gg mL^2/2\hbar$ so that the measured momentum distributions are unaffected by the initial spatial distribution of the atoms.

\section{Measurement of the contact}

\begin{figure*}[ht]
	\centering
	\includegraphics[width=1\textwidth]{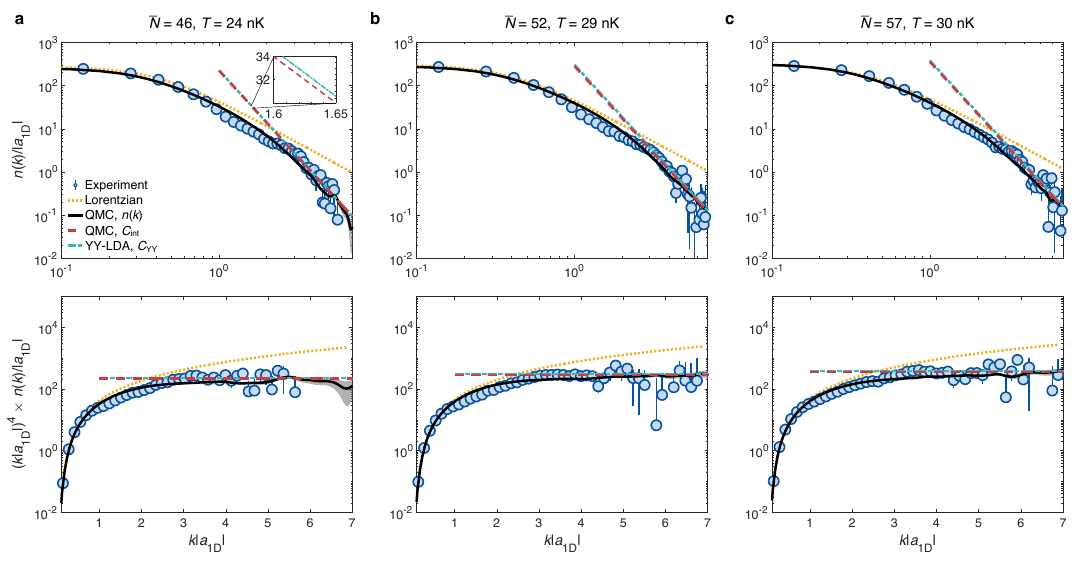}
	\caption{
		\textbf{Momentum distributions for the trapped Lieb-Liniger gas.}
		The upper row shows momentum distributions $n(k)$ in log-log scale for various average atom numbers $\overline{N}$ and temperatures $T$, indicated on top of each column. The lower row shows the corresponding quantities $n(k) \times k^4$ in semi-log scale for the same data.
		The figures display the experimental data (blue disks, with errorbars corresponding to the standard deviation), QMC results for the full momentum distribution (solid black line), Lorentzian fits to the low-$k$ sector (dotted orange line), as well as the QMC estimates using the thermodynamic Tan relation~(\ref{eq:C Hint}) (dashed red line, $C_{\textrm{int}}$) and the YY-LDA estimates using the Tan sweep relation~(\ref{eq:Csweep}) (dotted-dashed cyan line, $C_{\textrm{YY}}$) estimates. Note that the $C_{\textrm{int}}$ and $C_{\textrm{YY}}$ are on top of each other and hardly ditinguishable. The inset shows a magnified view of $C_{\textrm{int}}$ and $C_{\textrm{YY}}$.
	}\label{fig2}
\end{figure*}

Figure~\ref{fig2} shows three typical 1D momentum distributions, measured as discussed above, for various weighted average particle numbers $\overline{N}$ and temperatures $T$ (indicated on top of the figure). The upper row shows the momentum distribution $n(k)$ in log-log scale while the lower row shows $n(k) \times k^4$ in semi-log scale for the same data. In the experiment, the total atom number is found by integrating the full measured momentum distribution and the average atom number per tube, $\overline{N}$, is found as the weighted average over the tubes~\cite{NoteSuppl}. The temperature $T$ is estimated by comparison to QMC calculations~\cite{guo2024_1,guo2024_2} for the atom number $\overline{N}$ in a harmonic trap with frequency $\omega_x$ and various temperatures~\cite{NoteSuppl}. We estimate that of the experiment from the one that best fits the experimental data restricted to the low-$k$ sector (typically $k \lesssim 1/\vert\aOneD\vert$). We then obtain good agreement between the experimental data (blue dots) and the QMC calculations (solid black line), not only in the fitted low-$k$ sector but also in the tails with $k \gtrsim 1/\vert\aOneD\vert$, see Fig.~\ref{fig2}. In the low-$k$ sector, both experimental and QMC results are consistent with a Lorentzian momentum distribution (dotted orange line) with half width at half maximum (HWHM)
$k_{\phi}=\alpha \kB T/\hbar n_0$ with $T$ the fitted temperature and $n_0$ the atom density at the center of the trap. The heuristic parameter $\alpha$ encapsulates the effect of axial trapping and finite interactions~\cite{gerbier2003,richard2003,fabbri2011} and we found  $\alpha\approx 0.8$ for intermediate Lieb-Liniger parameter, $\gamma\sim 1$, relevant for our system~\cite{NoteSuppl}.

We now focus on the large-$k$ tails of the momentum distributions (typically $k \gtrsim 4/\vert\aOneD\vert$). There, the Lorentzian distribution breaks down and the data show algebraic tails, consistent with the expected universal behavior in $1/k^4$. This is visible in the lower row of Fig.~\ref{fig2}, which shows that the quantity $n(k)\times k^4$ approaches an asymptotic constant value. The latter is nothing but the Tan contact $C$. The experimental data for $n(k)$ (blue dots) are then compared to three independent theoretical estimates. First, the comparison to the QMC momentum distribution (solid black line) confirms the good agreement with the experimental data in semi-log scale.
Second, we directly compute the Tan contact from QMC calculations using the thermodynamic Tan relation~\cite{tan2008_2}
\begin{equation}\label{eq:C Hint}
	C_{\textrm{int}}=\frac{2\gOneD m^2}{\hbar^4}\langle H_{\textrm{int}} \rangle,
\end{equation}
where $\langle H_{\textrm{int}} \rangle$ is the average interaction energy found numerically (dashed red line). Note that the calculation of the latter is independent of the momentum distribution also calculated in the QMC. We nevertheless find good agreement between $C_{\textrm{int}}$ and the weight of the momentum tails found experimentally or in the QMC calculations. This confirms that the tails of the momentum distribution yield the Tan contact, which is thus accurately measured in the experiment. To our knowledge, this is the first direct measurement of the Tan contact for an interacting Bose gas from the momentum distribution. Third, we also compare the experimental result to the contact computed using the Tan sweep relation~\cite{tan2008_3}, 
\begin{equation}\label{eq:Csweep}
	C_{\textrm{YY}}=\frac{4m}{\hbar^2}\left.\frac{\partial \Omega}{\partial\aOneD}\right\vert_{T,\mu},
\end{equation}
where $\Omega$ is the grand potential, here calculated using Yang-Yang (YY) thermodynamics within local density approximation (LDA)~\cite{yao2018}.
We find fair agreement with both experimental data and QMC results. Such consistency between all estimates is found for all the experiments presented here, corresponding to the temperature range $20\,\textrm{nK}<T<40\,\textrm{nK}$ and average atom number per tube $34<\overline{N}<66$. More precisely, we find that the QMC estimate $C_{\textrm{int}}$ fits the experimental data  better than the YY-LDA estimate $C_{\textrm{YY}}$, which we attribute to the limited accuracy of LDA for the very low particle density achieved here.

\section{Thermodynamic properties of the Tan contact}

\begin{figure*}[ht]
    \centering
	\includegraphics[width=12cm]{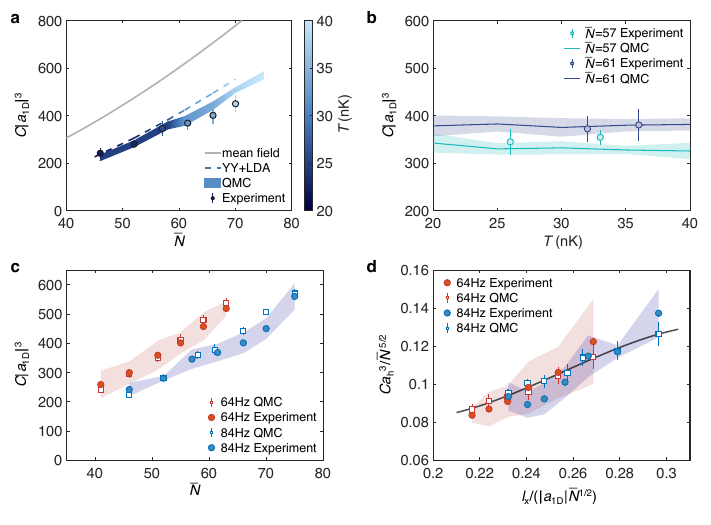}
	\caption{
		\textbf{Tan's contact versus atom number and temperature, and universal scaling.}
		\textbf{a},~Contact versus particle number $\overline{N}$. The figure shows the experimental data (disks with error bars, where the shades of blue represent the temperature), the QMC results for $C_{\textrm{int}}$ (solid areas represent error bars, equation~\ref{eq:C Hint}), the YY-LDA prediction (dashed blue line, equation~\ref{eq:Csweep}), and the mean-field prediction (solid gray line, equation~\ref{eq:C mean field}).
		\textbf{b},~Contact versus temperature for $\overline{N} = 57$ and $61$. The disks with error bars represent experimental data, while the solid lines denote QMC results, with shaded areas indicating the corresponding QMC error bars. Within a specific temperature range, both datasets indicate that $C$ remains nearly constant with temperature.
		\textbf{c},~Contact versus particle number for two sets of data corresponding to $\omega_x/2\pi=64\,\rm{Hz}$ (lattice amplitude $V=40\Er$, red markers) and $\omega_x/2\pi=84\,\rm{Hz}$ ($V=70\Er$, blue markers). The squares are QMC results while the disks are experimental data, and the shaded areas correspond to the experimental error bars.
		\textbf{d},~Same data as in c but plotted according to the scaling of equation~\ref{eq:scaling C}. The solid line is a fit to the QMC data.
	}\label{fig3}
\end{figure*}

In the experiment, the Tan contact $C$ is extracted from fits to the asymptotic, large $k$, limit of the data for $n(k)\times k^4$ as in the lower row of Fig.~\ref{fig2}. The results for various values of the weighted average particle number $\overline{N}$ are shown in Fig.~\ref{fig3}a, where the particle number is tuned by controlling the duration of the evaporative cooling process.
At the same time, the final temperature $T$ (encoded in the color scale) varies, typically between $25$\,nK and $35$\,nK. However, it is expected that in the considered regime $\vert\aOneD\vert/\lambda_{\textrm{dB}} \sim0.5$, with $\lambda_{\textrm{dB}}=\sqrt{2 \pi \hbar^2/m\kB T} \approx 1.08\times 10^{-6}$\,m the de Broglie wavelength at $T=30\,\mathrm{nK}$ and $\vert\aOneD\vert \approx 5.8 \times 10^{-7}$\,m, the value of the contact $C$ does not significantly depend on the temperature~\cite{kheruntsyan2003,yao2018}. To check this, we take advantage of the temperature fluctuations and extract, from our set of measurements, data with a fixed average number of particles but various temperatures. A representative result for $\overline{N}=57\pm 2$ is shown in Fig.~\ref{fig3}b. It confirms that, in the temperature range we achieved here, the value of $C$ is almost constant within about $10\%$ and compatible with the theoretical prediction (solid line). In the following, we can thus disregard the effect of temperature fluctuations on the contact $C$.

Consistently with theoretical predictions, we find that the contact increases monotonically with the particle number, see Fig.~\ref{fig3}a. To understand the experimental data, we may first compare them with the mean-field prediction~\cite{olshanii2003,kheruntsyan2003,yao2018}
\begin{equation}\label{eq:C mean field}
	C = \frac{\eta N^{5/3}}{\lx^{4/3} \aOneD^{5/3}},
\end{equation}
where $\eta = 4\times 3^{2/3}/5 \simeq 1.66$ and $\lx = \sqrt{\hbar/m\omega_x}$ is the axial harmonic oscillator natural length, shown as a solid gray line in Fig.~\ref{fig3}a. We then find that our data strongly deviate from this simple estimate by more than $46\%$ for all data points. We many now compare the experimental data to many-body calculations based on both QMC, equation~\ref{eq:C Hint} (solid blue line), and YY-LDA (dashed blue line). Clearly, the experimental data is in excellent agreement with the exact QMC calculations, within $12\%$. The YY-LDA is also in fair agreement with the experimental data, although with a larger deviation of about $21\%$.
The fact that the experimental data are consistent with exact many-body predictions but significantly differ from a mean field calculation indicates strong beyond mean-field effects. This is consistent with the fact that the Lieb-Liniger parameter $\gamma$ ranges from 1.0 to 1.4 in the experiment.

\section{Universal scaling of the Tan contact}

Finally, our experimental setup allows us to test the predicted universal two-parameter scaling of the Tan contact,
\begin{equation}\label{eq:scaling C}
	C = \frac{N^{5/2}}{\lx^{3}}  f \left(\xig, \xiT \right),
\end{equation}
with the interaction parameter $\xig=-\lx/\aOneD\sqrt{N}$ and the temperature parameter $\xiT=-\aOneD/\lambda_{\textrm{dB}}$~\cite{yao2018}.  By varying the 2D lattice amplitude creating the tubes, we can tune the longitudinal trapping frequency $\omega_x$ and the coupling constant $\gOneD$, or equivalently the parameters $\lx$ and $\aOneD$. We choose two values of the lattice potential depth: (i)~$V=40\,\Er$, which corresponds to $\omega_x/2\pi \approx 64$\,Hz and $\aOneD \simeq -0.2\,\mu\rm{m}$, and (ii)~$V=70\,\Er$, which corresponds to $\omega_x/2\pi \approx 84$\,Hz and  $\aOneD \simeq -0.15\,\mu\rm{m}$. Plotting $C \lvert \aOneD \rvert^3$ versus the particle number $\overline{N}$, as in Fig.~\ref{fig3}c, we observe two clearly separated groups of data, corresponding to each set of parameters ($V=40\,\Er$, red markers; $V=70\,\Er$, blue markers). Again, the experimental (disks) and QMC (hollow squares) results are in good agreement. Then, plotting the same data using the rescaled quantities $\tilde{C}= C\lx^{3}/N^{5/2}$ and $\xig=-\lx/\aOneD\sqrt{N}$, we observe data collapse within errorbars, see Fig.~\ref{fig3}d. This is consistent with the universal two-parameter scaling of equation~\ref{eq:scaling C}. The solid black line in Fig.~\ref{fig3}d is a fit to the QMC data and represents the universal scaling function $f$ versus $\xig$.
Since the temperature variation is negligible in the considered experimental conditions, the value of $\xiT$ is fixed.

\section{Conclusions}

Our work reports the first experimental measurement of the Tan contact $C$ in a correlated Lieb-Liniger gas, from the direct observation of the large-momentum tails. Leveraging the quasi-1D structure of our system, we devised a two-stage TOF imaging sequence, which prevents detrimental effects of interactions during the gas expansion. Comparison of the observed short-momentum distribution with QMC calculations allows us to determine the experimental temperature while the weight of the large-momentum tails yields a direct measure of $C$. Very good agreement is found with theoretical predictions and we show clear beyond mean-field effects. Varying the particle number, the temperature, as well as the interaction strength via transverse confinement, allowed us to realize a first test of the universal two-parameter scaling law predicted in Ref~\cite{yao2018}.

Direct measurement of the Tan contact as realized here provides a wealth of information about the thermodynamics of the Lieb-Liniger gas. The accessible ranges of atom number and temperature is presently limited in our system but this can be improved by adapting the laser cooling and evaporative stages of the atomic gas preparation. Using stronger transverse confinement can also be used to access stronger interaction regimes. It would allow for systematic demonstration of the two-parameter scaling law. It would for instance pave the way to observation of fermionization of the strongly-interacting Lieb-Liniger gas at finite temperatures, which is marked by a characteristic maximum of the rescaled Tan contact~\cite{yao2018}. Our approach may also be extended to further quantum models in confined geometries, for instance ultracold Fermi gases and Bose-Bose, Bose-Fermi or Fermi-Fermi mixtures, the thermodynamics of which may also be characterized by the properties of the contact~\cite{bloch2008,sagi2012,decamp2016,patu2018,zou2021}. While Feshbach resonances may be alternatively used to switch off interactions during time-of-flight imaging in single-component gases, our approach has the advantage of being applicable also to mixtures.

\acknowledgments
We thank David~Cl\'ement for useful comments on the manuscript.
This research was supported by the Agence Nationale de la Recherche (ANR, project ANR-CMAQ-002 France~2030) and the program ``Investissements d'Avenir'', the LabEx PALM (project ANR-10-LABX-0039-PALM), the Swiss National Science Foundation under grant numbers 200020-188687 and 200020-219400, and the National Natural Science Foundation of China (Grants Nos. 11920101004).

\bibliography{bibliography}

\renewcommand{\theequation}{S\arabic{equation}}
\setcounter{equation}{0}
\renewcommand{\thefigure}{S\arabic{figure}}
\setcounter{figure}{0}
\renewcommand{\thesection}{S\arabic{section}}
\setcounter{section}{0}
\onecolumngrid  

\newpage

{\center \bf Supplemental Material for \\}
{\center \bf \large Direct measurement of Tan's contact in a one-dimensional Lieb-Liniger gas \\ \vspace*{1.cm}
}

\section{Initial transverse dynamics in the two-stage expansion scheme}\label{App:criteria}

The transverse dynamics in the first stage of our two-stage expansion scheme is that of a weakly-interacting 2D Bose gas, owing to the strong initial transverse confinement. It is governed by the time-dependent Gross-Pitaevskii equation
\begin{equation}\label{eq:TransverseSchrodinger}
    i \hbar \frac{\partial \psi(\mathbf{r},t)}{\partial t} = \left[ \frac{-\hbar^2 \mathbf{\nabla}^2}{2m} + \frac{m\omegap^2(t)}{2}\mathbf{r}^2 + g \left\vert \psi(\mathbf{r},t)\right\vert^2\right]\psi(\mathbf{r},t),
\end{equation}
where $g$ is the 2D coupling constant and $\psi(\mathbf{r},t)$ is the transverse wavefunction with $\mathbf{r}=(y,z)$. For an arbitrary time-evolving trap frequency $\omegap(t)$, the solution is self-similar and the atom density $n(\mathbf{r},t)=\left\vert\psi(\mathbf{r},t)\right\vert^2$ fulfills $n(\mathbf{r},t)=n(\mathbf{r}/b(t),0)$~\cite{castin1996,kagan1996}. Note that while this is correct only in the Thomas-Fermi regime in 3D, it is universal, that is irrespective of the interaction strength, in 2D~\cite{kagan1996}. The expansion factor $b(t)$ is then governed by the second-order differential equation
\begin{equation}\label{eq:expansion-normal}
    \ddot{b}(t)+\omegap^2(t) b(t) = \frac{\omegap^2(0)}{b^3(t)}.
\end{equation}
In the experiment, the transverse confinement is created by a 2D optical lattice of amplitude $V(t)$, and we have $\omegap(t)=\hbar k^2 \sqrt{V(t)/\Er}/m$.

\begin{figure}[ht]
	\centering
	\includegraphics[width=8cm]{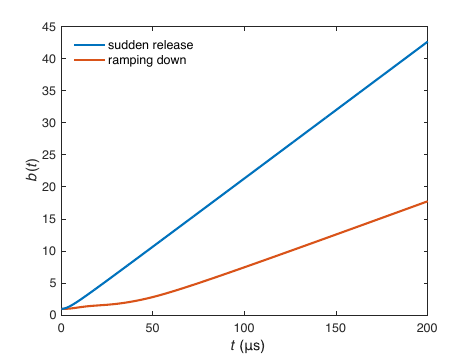}
	\caption{
        Time evolution of the expansion parameter $b(t)$ for a sudden release and the exponential ramp.
	}\label{bt evolution}
\end{figure}

For a sudden switch-off of the 2D optical lattice, $\omegap(t)=0$ for any $t>0$, and Eq.~(\ref{eq:expansion-normal}) may be solve exactly~\cite{castin1996},
\begin{equation}\label{eq:expansion quench}
    b(t)=\sqrt{1+[\omegap(0) t]^2}.
\end{equation}
In the two-stage expansion scheme, we switch off the 2D optical lattice in a time $\Delta t$ following an exponential ramp such that
\begin{equation}\label{eq:RampTOF}
    \omegap(t)= \omegap(0) \left(\frac{\exp(-4t/\Delta t+4)-1}{\exp(4)-1}\right)^{1/2}
\end{equation}
for $0<t<\Delta t$ and $\omegap(t)=0$ for $t \geqslant \Delta t$. Equation~(\ref{eq:expansion-normal}) is then solved numerically. The result is shown in Fig.~\ref{bt evolution} and compared to Eq.~(\ref{eq:expansion quench}), for $\Delta t=50\,\mu\rm{s}$ and a total expansion time $t_{\rm{max}}=1/10\omega_x = 200\,\mu\rm{s}$.
We find that the exponential ramp significantly slows down the transverse expansion and we find $b(1/10\omega_x) \simeq 17.8$.

\section{\quad Weighted average atom number and temperature estimate}\label{app:TandN}

Here, we discuss the estimates of the atom number distribution and and temperature of our system, as well as QMC simulations within a single tube with weighted average atom number.

\subsection{Atom number distribution}

In the beginning of the experiment, we obtain a 3D BEC in a nearly harmonic trap with frequencies
$\Omega_x/2\pi \approx 23\,\mathrm{Hz},\Omega_y/2\pi \approx 20\,\mathrm{Hz},\Omega_z/2\pi \approx 28\,\mathrm{Hz}$.
For the large particle number in our experiment, we are in the Thomas-Fermi regime where the 3D Gross-Pitaevskii equation for the BEC wave function $\psi(\mathbf{r})$ reduces to
\begin{equation}
	\left[ V(\mathbf{r})+g \lvert \psi(\mathbf{r}) \rvert ^{2} \right] \psi(\mathbf{r})=\mu \psi(\mathbf{r}),
\end{equation}
where $V(\mathbf{r})$ is the harmonica trap potential,
$g$ the 3D coupling constant, and $\mu$ the chemical potential.

The 2D lattice is then ramped up, which cuts the system into 1D tubes. Since the loading process is slow, we can compute the atom number distribution using the rescaled coupling constant~\cite{kramer2002}
\begin{equation}
	\tilde{g}=g\frac{\pi (V_0/\Er)^{1/2}}{2 \left( \operatorname{Erf}\left[\pi (V_0/\Er)^{1 / 4} / 2\right]\right)^2},
\end{equation}
where $V_0$ is the 2D lattice depth. Such a description includes the effect of the system expansion during the loading of lattice. The corresponding chemical potential is
\begin{equation}
	\mu^{\prime}=\frac{\hbar \bar{\omega}}{2}\left(15 N \frac{a_{\mathrm s}}{\bar{\ell}} \frac{\tilde{g}}{g}\right)^{2 / 5},
\end{equation}
where $\bar{\ell} = \sqrt{\hbar/m\bar{\omega}}$
with $\bar{\omega}=\sqrt{\omega_x \Omega'_y \Omega'_z}$
is the oscillation length and $a_{\textrm{s}}$ is 3D s-wave scattering length. Note that the transverse trapping frequencies are slightly modified due to the Gaussian transverse shape of the lattice beams and are nearly equal, $\Omega'_y \approx \Omega'_z \approx \Omega'_\perp$.
With these two values, we can estimate the Thomas-Fermi radius in the transverse directions
\begin{equation}
	\RTF=\sqrt{\frac{2 \mu^{\prime}}{m \Omega_{\perp}^{\prime 2}}},
\end{equation}
and the atom number distribution,
\begin{equation}\label{eq:Nij}
	N_{i, j}=N_{0,0}\left[1-\frac{i^{2}+j^{2}}{(\RTF/d)^{2}}\right]^{3/2},
\end{equation}
where $i,j$ is the index number along $y,z$ directions, $d$ is the lattice spacing, and $N_{0,0}$ is the number of atoms in the central tube.  The value of $N_{0,0}$ can be deduced from the total atom number $N$ with the condition $N =\sum_{i,j}  N_{i,j}$, which yields
\begin{equation}
	N_{0,0} = \frac{5N d^2}{2 \pi \RTF^{2}},
\end{equation}
and the weighted average atom number is then
\begin{equation}\label{eq:Naverage}
	\overline{N}=\sum_{i,j} \frac{N_{i,j}^2}{N}.
\end{equation}
The Lieb-Liniger parameter is then estimated according to the averaged density $\bar{n}=\bar{N}/L_{\bar{N}}$ with $L_{\bar{N}}$ the tube length computed from the QMC calculations, $\gamma = m\gOneD/\hbar^2\bar{n}$.

\subsection{Single tube QMC calculations with weighted average atom number}

\begin{figure}[ht]
	\centering
	\includegraphics[width=8cm]{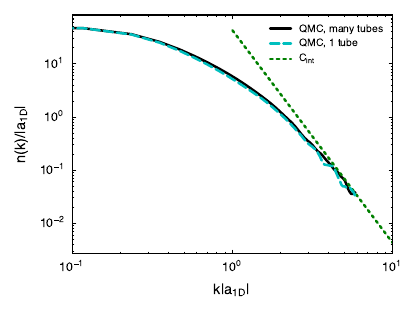}
	\caption{
        Comparison of momentum distributions $n(k)$ computed with QMC for the full tube distribution (black solid line) and one single tube with weighted average atom number (blue dashed line), for the same parameters as in Fig.~\ref{fig2}b of the main text. The green dashed line shows the $k^{-4}$ tail predicted by the Tan sweep relation.
    }\label{sfig-tube}
\end{figure}

When comparing our experimental data with quantum Monte Carlo (QMC) calculations, we always simulate our system by one single tube with  a number of atoms equal to the weighted average atom number $\overline{N}$ as estimated above, instead of simulating all tubes. Here, we check this approximation and show that is very accurate. Figure~\ref{sfig-tube} shows a comparison of momentum distributions for the case $\overline{N}=52$ and $T=23nK$, corresponding to Fig.~\ref{fig2}b of the main text, where we simulate all tubes with the atom distribution of equation~\ref{eq:Nij} (black solid line) or one single tube with the weighted average number of atoms of equation~\ref{eq:Naverage} (blue dashed line). Clearly, they show very good agreement for both small and large $k$ regimes. In particular, they show the same $k^{-4}$ tail with weights (contact $C$) differing by less $8\%$.
The sweep relation prediction is also shown (green dashed line) for reference.

\subsection{Temperature}

\begin{figure}[ht]
	\centering
	\includegraphics[width=8cm]{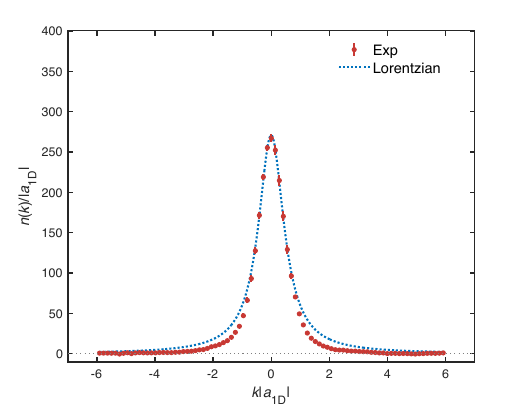}
	\caption{
        Typical example of Lorentzian fit to the momentum distribution $n(k)$ found in the experiment for a weighted average atom number of $\overline{N}\approx 57$ and trapping frequency $\omega_x/2\pi \approx 84$.
    }\label{fig_s3}
\end{figure}

The temperature in the experiment is estimated by a Lorentzian fit of the low-$k$ sector of the momentum distributions. The Lorentzian shape is expected for a weakly-interacting, homogeneous 1D Bose gas~\cite{kane1967}. In the presence of a harmonic trap, it has been found that the Lorentzian shape still holds for weakly-interacting quasi-condensates with the momentum HWHM, 
\begin{equation}\label{eq:DeltaP}
	\Delta p = \frac{\alpha}{L_\phi},\ L_\phi = \frac{\hbar^2 n_0}{m k_{\textrm{B}} T}
\end{equation}
where $L_\phi$ is the thermal length, $n_0$ is the central particle density, and $\alpha$ is a phenomenological correction factor, which depends on the interaction strength~\cite{gerbier2003,richard2003,fabbri2011}.
Here we find that the same holds for the Lieb-Liniger gas with stronger interactions and we find a parameter $\alpha \simeq 0.8$ in the QMC simulations. Figure\,\ref{fig_s3} shows an example of a fit of a Lorentzian function to the 1D momentum distribution found in the experiment, the HWHM of which yields the temperature of the experiment through equation~\ref{eq:DeltaP}.

\end{document}